\documentclass[a4paper,fleqn,usenatbib]{mnras}

\usepackage[T1]{fontenc}
\usepackage{ae,aecompl}
\usepackage{hyperref}
\usepackage{graphicx}	% Including figure files
\usepackage{amsmath}	% Advanced maths commands
\usepackage{amssymb}	% Extra maths symbols

\title[Discovery of a GeV extended source]{Discovery of an extended source of gamma-ray emission in the Southern Hemisphere}

\author[Araya]{
Miguel Araya$^{1}$\thanks{E-mail: miguel.araya@ucr.ac.cr} 
\\
$^{1}$Centro de Investigaciones Espaciales and Escuela de F\'isica, Universidad de Costa Rica
}

\date{Accepted XXX. Received YYY; in original form ZZZ}

\pubyear{2017}

\begin{document}
\label{firstpage}
\pagerange{\pageref{firstpage}--\pageref{lastpage}}
\maketitle

\begin{abstract}
We report the discovery of a $\sim$ $3\degr\llap{.}4$-wide region of high-energy emission in data from the \emph{Fermi} LAT satellite. The centroid of the emission is located in the Southern Hemisphere sky, a few degrees away from the plane of the Galaxy at the Galactic coordinates l=350$\degr\llap{.}$6, b=-4$\degr\llap{.}$7. It shows a hard spectrum that is compatible with a simple power-law, $\frac{dN}{dE}\propto E^{-\Gamma}$, in the energy range 0.7--500 GeV, with a spectral index $\Gamma = 1.68 \pm 0.04_{\mbox{\tiny stat}} \pm 0.1_{\mbox{\tiny sys}}$. The integrated source photon flux above 0.7 GeV is $(4.71 \pm 0.49_{\mbox{\tiny stat}}\pm 2.13_{\mbox{\tiny sys}}) \times 10^{-9}$ cm$^{-2}$ s$^{-1}$. We discuss several hypotheses for the nature of the source, particularly that the emission comes from the shell of an unknown supernova remnant.
\end{abstract}
%instruction for authors:http://www.oxfordjournals.org/our_journals/mnras/for_authors/

\section{Introduction}\label{sec:intro}
The current generation of gamma-ray observatories have revealed a large variety of Galactic and extragalactic sources of high-energy (MeV to GeV) and very high-energy (TeV) emission. In particular the all-sky survey of the \emph{Fermi} Large Area Telescope (LAT) \citep{2009ApJ...697.1071A} has had considerable impact on the science of high-energy astrophysics. Gamma-ray emitting objects observed by the LAT include supernova remnants (SNRs), and their surroundings, pulsar wind nebulae (PWN), pulsars, binary systems, novae, and supermassive black holes \citep[e.g.,][]{2010ApJ...710L..92A,2013ApJ...773...77A,2010ApJS..187..460A,2009ApJ...701L.123A,2014Sci...345..554A,2015ApJ...810...14A}.

Uncovering the origin of gamma-ray emission in all these systems is important to understand the physical processes behind the acceleration of particles. Its study may also help reveal the origin of Galactic cosmic rays with energies up to $\sim$PeV. Several SNRs have been established as sources of high-energy cosmic rays \citep[e.g.,][]{2013Sci...339..807A} but the search for the Galactic PeVatron continues and many of the observed sources remain unidentified.

SNRs also show a variety of gamma-ray spectra and features that are not fully understood, which points to the need for more studies and theoretical work on the problem of particle acceleration and transport \citep[e.g.,][]{2009ApJ...706L...1A,2011ApJ...735..120Y,2012JCAP...07..038C,2014BrJPh..44..415B,2016ApJS..224....8A}.

Many known Galactic extended sources at GeV and TeV energies are likely PWN or SNRs \citep{2015ApJS..218...23A,2013arXiv1307.4690C}. Well studied examples of SNR shells that emit gamma-rays are RX J1713.7-3946 \citep{2011ApJ...734...28A} and RX J0852.0-4622 \citep{2011ApJ...740L..51T}. These sources show a hard GeV spectrum which can be explained by inverse Compton (IC) scattering of soft photons by high-energy leptons, although the origin of the emission is still debated. Most gamma-ray sources are found close to the Galactic Equator, but this is not always the case. A notable exception is the unidentified source HESS J1507-622 \citep{2011A&A...525A..45H} that lies about 3$\degr\llap{.}$5 from the plane of the Galaxy and has no clear X-ray counterpart, which is surprising given the low absorption expected at its location. This also prevents a clear determination of its distance, a key parameter to understand its nature. Some authors have suggested that this and other unidentified gamma-ray sources could be ancient PWN with no counterpart at other wavelengths \citep{2013ApJ...773..139V}.

Here, we report the discovery of an extended source of gamma-rays seen in LAT data, which shows some of the features of the sources discussed above. The source shows a hard photon spectrum. When modeled with a uniform disc template, the center of the emission is located at the Galactic coordinates $l=350\degr\llap{.}6$, $b=-4\degr\llap{.}7$ and its best-fit radius is $\sim 1\degr\llap{.}7$. Previously, several gamma-ray sources were found in the region as well as hints of the presence of extended high-energy emission. The Third EGRET Catalog of High-Energy Gamma-Ray Sources \citep{1999ApJS..123...79H} shows a relatively low-significance unidentified source, 3EG J1744-3934 which can be consistent with an extended source and also with a collection of multiple sources. It is located at the Galactic coordinates $l=350\degr\llap{.}81$, $b=-5\degr\llap{.}38$. A power-law spectral index of $\sim 2.4$ is reported for this EGRET source. In the \emph{Fermi} LAT 4-year Point Source Catalog \citep[3FGL,][]{2015ApJS..218...23A}, the unidentified sources 3FGL J1733.5-3941 and 3FGL J1748.5-3912 lie close to the boundary of the new source reported here and show soft spectra. The source 3FGL J1747.6-4037 is also near the boundary of the new extended source. It is identified as the gamma-ray counterpart of the millisecond pulsar PSR J1747-4036 with a dispersion measure that implies a source distance of 3.4 kpc. The pulsar's spindown luminosity is $1.16\times 10^{35}$ erg s$^{-1}$ \citep[uncorrected for the Shklovskii effect;][]{2012ApJ...748L...2K,2013ApJS..208...17A}. It is a faint gamma-ray source and the LAT's spectrum shows no evidence of spectral curvature.

A $\sim31\arcmin$-wide SNR, G351.0-5.4, was recently discovered in radio observations of the region \citep{2014A&A...568A.107D}. The SNR has a low surface radio brightness, its centre is located at the Galactic coordinates $l=351\degr\llap{.}06$, $b=-5\degr\llap{.}49$ and has no optical counterpart, which led the authors to classify it as an old SNR. No PWN is known within the extent of the region. \cite{2014A&A...568A.107D} analyzed Pass 7 LAT data around the SNR and found a slightly spatially extended feature above the background with a 1.5$\sigma$ confidence, although no hints of emission within the radio contours of the SNR were found. Finally, the unidentified point-like sources 2FHL J1741.2-4021, reported in the Second \emph{Fermi} LAT Catalog of High-Energy Sources (2FHL) detected above 50 GeV \citep{0067-0049-222-1-5}, is found within the region, and 3FHL J1733.4-3942 reported in the Third Catalog of Hard \emph{Fermi} LAT Sources \citep[3FHL,][]{0067-0049-232-2-18}, was seen at the edge of the new gamma-ray source reported here. In this paper LAT data selection, analysis and results are described in Section \ref{sec:data}. In Section \ref{sec:discussion} possible interpretations of the nature of this source are given.

\section{Data analysis}\label{sec:data}
\emph{Fermi} LAT data were gathered from 2008 August to 2016 September from the LAT data server\footnote{See http://fermi.gsfc.nasa.gov}. The reprocessed PASS 8 photon and spacecraft data were used. The instrument response functions P8R2\_SOURCE\_V6 were used and the data were analyzed with the public Science Tools software version v10r0p5. Recommended cuts for standard analysis were applied by selecting SOURCE class events and a maximum zenith angle selection of 90$\degr$ to avoid albedo gamma rays from the Earth, time intervals when the data quality was good and the data taking mode was standard.

A 20$\degr\times20\degr$ region of interest centered at the Galactic coordinates $l=350\degr\llap{.}$33, $b=-4\degr\llap{.}$46 was used in the analysis with events having energies between 200 MeV and 500 GeV. The data were binned spatially in counts maps with a scale of 0$\degr\llap{.}$05 per pixel and in energy with ten logarithmically spaced bins per decade for exposure calculation.

The spectral and morphological properties of sources in the region were studied with a maximum likelihood optimization \citep{1996ApJ...461..396M} including all the sources in the 3FGL catalog within the region of interest, the standard galactic diffuse emission model \citep[described by the file gll\_iem\_v06.fits,][]{2016ApJS..223...26A} and the isotropic component (iso\_P8R2\_SOURCE\_V6\_v06.txt) distributed with the LAT analysis software. During all the fits the normalizations of the diffuse components were left free as well as the normalizations of the sources located within 5$\degr$ of the center of the region of interest. For one additional degree of freedom in the fit, the significance of detection of a source can be estimated as the square root of the test statistic (TS), defined as $-2\times$log$(L_0/L)$, where $L_0$ is the likelihood value without the source and $L$ the likelihood with the additional source.

\subsection{Morphology of the emission}\label{subsec:morph}
\begin{figure}
    \includegraphics[width=\columnwidth]{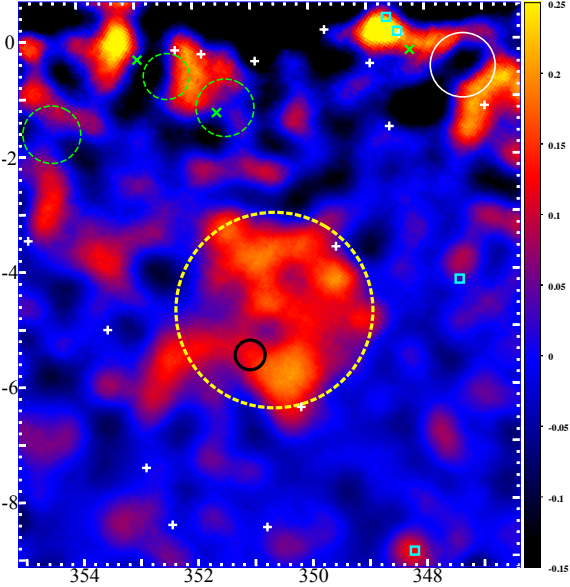}
    \caption{LAT counts map above 5 GeV obtained after subtraction of the final background model using a scale of $0\degr\llap{.}05$/pixel. The image is smoothed with a Gaussian kernel with $\sigma=0\degr\llap{.}$3. The yellow circle represents the best fit disc template found for G350.6-4.7 and the black circle the radio extent of SNR G351.0-5.4 \citep{2014A&A...568A.107D}. The white crosses (circle) indicate the positions of point like (extended) sources in the background model from the 3FGL catalog, the green circles and X's represent the new background sources found in this work and the magenta boxes are the sources from the 3FHL catalog that were included in the model. The vertical and horizontal axes are the Galactic latitude and longitude in degrees, respectively.}
    \label{fig1}
\end{figure}

Only events above 5 GeV were used to study the morphology of the emission in order to take advantage of the narrower LAT Point Spread Function at higher energies. From the 3FGL sources close to the center of the region of interest (see Fig. \ref{fig1}), only the source 3FGL J1733.5-3941 was significantly detected (with a significance of 6.7$\sigma$) and in order to facilitate convergence in the following its spectral parameters were fixed at the best fit values. This source showed a soft spectrum that was different to the spectrum of the new source found here and thus it was left in the model. The other sources in the region are 3FGL J1748.5-3912, which has no association in the 3FGL catalog and it is thus removed from the model as it could be part of the emission from the new source, and 3FGL J1747.6-4037, which is left in the model as it is associated to the pulsar PSR J1747-4036.

After the initial fit a map of the background was created and subtracted from the observed counts map. In order to lower the effect of residual background from unmodeled emission in the Galactic plane, several disc templates were used to look for extended emissions. Their positions, radii and spectral parameters assuming a simple power law shape were optimised by searching for the maximum TS values around a starting location. The templates added were two 0.5$\degr$-radius discs centered at $(l,b)=$ 351$\degr\llap{.}$44, -1$\degr\llap{.}$18 and 354$\degr\llap{.}$44, -1$\degr\llap{.}$61; and a 0.4$\degr$-radius disc centered at 352$\degr\llap{.}$45, -0$\degr\llap{.}$626. Additionally, the point sources found in the 3FHL catalog \citep{0067-0049-232-2-18} that are located within the region of interest were added to the model to improve the description of the emission (with the exception of 3FHL\_J1733.4-3942 which is located at the edge of the extended region of gamma-ray emission found here). A new search for point sources was carried out along the Galactic plane which resulted in three new sources with a TS above 25, which were also included in the model.

The final map of residual emission obtained after subtracting the new background sources described above as well as the sources in the 3FGL and 3FHL catalogs is shown in Fig. \ref{fig1} in Galactic coordinates. Enhanced emission from a region extending several degrees around the center of the image is evident. In order to study the morphology of this emission, the region is modeled with two spatial hypotheses: a uniform disc and a collection of point sources. Due to low statistics and to facilitate convergence, the spectra used in all cases was a simple power law. The disc template was placed within the interior of the region and its center position moved in a $4\degr\times4\degr$ square grid with the origin at the center of the region of interest, in steps of 0$\degr\llap{.}$1, changing its radius systematically at each position in steps of 0$\degr\llap{.}$1 from 0$\degr\llap{.}$5 to 2$\degr\llap{.}$0, while at the same time fitting the spectral normalization and index to find the maximum likelihood.

A systematic search for point sources within the excess seen in Fig. \ref{fig1} was carried out with the optimization tool gtfindsrc, provided with the standard LAT analysis software. However, the highest significance found for a point source in the region was 4$\sigma$. When the likelihood value for a model containing the four point sources with the highest significances in the region is compared to the null hypothesis, the resulting TS is 54. This model contains 16 additional degrees of freedom. The uniform disc model, with only 5 additional degrees of freedom and a TS of 144 with respect to the null hypothesis, is preferred. The best-fit template is centered at the Galactic coordinates $(l,b)=350\degr\llap{.}63,-4\degr\llap{.}71$ and has a radius of $1\degr\llap{.}7\pm 0\degr\llap{.}2$. The 3-$\sigma$ uncertainty in the position is $0\degr\llap{.}3$ and was estimated from the change in the source TS resulting from displacing the disc from the best-fit position. The new gamma-ray source is thus referred to as G350.6-4.7 from now on. As a cross check, a similar search for extended emission was done above 50 GeV where the diffuse emission intensity is much lower. The highest significance above 50 GeV for a disc template is $7.8\sigma$ and its corresponding size and location exactly match those found in this Section.

It is noted that a different spatial hypothesis containing several independent and extended sources in the region was not studied here, and this is left for future work. The same can be said about testing other spatial templates such as emission from a shell or probing for spatial substructure. It is also possible that the emission from G350.6-4.7 is the result of several extended objects (e.g., supernova remnants) present in this region and a more detailed analysis to disentangle possible distinct contributions is left for the future. However, the spectral fits used to find the disc template for G350.6-4.7 by placing the discs in different positions in a large region around the center of the source did not show variations in the spectral index, which might indicate that the emission comes from the same object.

\subsection{Spectrum}\label{subsec:spec}
The best-fit spatial template found in Section \ref{subsec:morph} and the improvements in the background model were used to obtain the source spectrum in the LAT energy range. At the lowest energies some marginally significant residual emission is found in the region having a very steep spectrum, which might not be related to the source. Given the estimated LAT differential sensitivity\footnote{https://www.slac.stanford.edu/exp/glast/groups/canda\\/lat\_Performance.htm}, if the hard spectrum of G350.6-4.7 extends below 1 GeV without a break, it is expected that the source would not be detected below energies of $\sim 0.7$ GeV, which is chosen as the energy threshold for the rest of the analysis. A more detailed study of the low energy spectrum of the source is left for the future.

In order to probe for curvarture in the spectrum different spectral shapes were used in independent likelihood fits above 0.7 GeV: a simple power law, a log parabola and a power-law with an exponential cutoff. The TS values obtained for each spectral assumption are, respectively, 217, 220 and 218. For the power law with exponential cut off the fit was not able to constrain the cutoff energy as it tried to find a value above the high energy limit of 500 GeV in the data. Both of the curved spectral hypotheses have one additional degree of freedom with respect to the simple power law, and for the highest TS value obtained with the log parabola function, the significance of curvature is $\sqrt{3}\sim 1.7\sigma$, which is not significant. These results imply that the best-fit spectral shape of this type for G350.6-4.7 is a simple power law ($\frac{dN}{dE}\propto E^{-\Gamma}$) in the LAT energy range.

The best-fit power-law index is $\Gamma = 1.68 \pm 0.04_{\mbox{\tiny stat}} \pm 0.1_{\mbox{\tiny sys}}$ and the integrated flux above 0.7 GeV is $(4.71 \pm 0.49_{\mbox{\tiny stat}}\pm 2.13_{\mbox{\tiny sys}}) \times 10^{-9}$ cm$^{-2}$ s$^{-1}$. The overall detection significance is 14.7$\sigma$. Since initial submission of this paper and dissemination of the work, preliminary results from the study of a new source has been reported at the International Cosmic Ray Conference (ICRC2017) by the \emph{Fermi} LAT Collaboration \citep{2017arXiv170906213W}, the source shows the same morphology as G350.6-4.7 and a similar spectrum which confirms our results.

\subsubsection{Spectral energy distribution}
In order to obtain the flux of the source in independent energy bins, the spectral energy distribution was constructed using ten energy intervals in the range 0.7--500 GeV. A likelihood fit was performed within each of the intervals with all parameters fixed for the sources in the model except for the normalization of the diffuse models and the normalization of the template representing the source G350.6-4.7. In all the energy bins, the source TS was greater than 5. The resulting spectral energy distribution values and their 1$\sigma$ statistical uncertainties are shown in Table \ref{tab2}.

\begin{table*}
	\centering
	\caption{Spectral energy distribution data points obtained for G350.6-4.7.}
	\label{tab2}
	\begin{tabular}{cc}
          \hline
	  \hline
	  Energy bins (GeV) & $\nu F_{\nu}$ ($10^{-12}$ erg cm$^{-2}$ s$^{-1}$) \\
          \hline
	   0.7--1.4	 & $4.91 \pm 1.05$\\
	   1.4--2.6     &  $4.17 \pm 1.21$\\
	   2.6--5.0      & $4.45 \pm 1.38 $ \\
           5.0--9.5      & $9.93 \pm 1.72$ \\
	   9.5--18	 & $6.09 \pm 1.97$ \\
	   18--35      & $8.04 \pm 2.42$ \\
	   35--66      & $10.8 \pm 3.25$ \\
           66--126      &  $16.0 \pm 4.44$\\
	   126--240	 & $10.1 \pm 5.24$ \\
	   240--500	 & $24.2 \pm 8.09$  \\
		\hline
	\end{tabular}\\
%%%%%%%\textsuperscript{a}\footnotesize{1$\sigma$ statistical uncertainty.}\\
\end{table*}

\subsection{Systematic errors}\label{subsec:sys}
An important systematic effect in the analysis of LAT data comes from the uncertainty in the model of the diffuse emission which is most important at energies below a few GeVs. Previous studies estimated that the intensity of diffuse gamma-rays can vary by $\pm 6$\% along the Galactic plane with respect to the standard model \citep{2009ApJ...706L...1A,2012ApJ...756...88C,2015A&A...574A.100H}. Regarding the uncertainty in the angular distribution of the diffuse emission, \cite{2015A&A...577A..12F} carried out a detailed study of the effect of the HI absorption uncertainty using high-resolution gas surveys in the region. They found that for the region of interest studied here, mostly located outside the Galactic plane, the difference between the standard LAT diffuse model and their refined model is negligible.

However, a more complete analysis of the effect of the uncertainties in the background emission model was carried out following \cite{2016ApJS..224....8A}, who used a set of eight alternative models\footnote{Found at https://fermi.gsfc.nasa.gov/ssc/data/access/lat\\/1st\_SNR\_catalog/} developed in \cite{2012ApJ...750....3A} by changing some crucial parameters that determine the intensity and distribution of the Galactic gamma-ray emission: the cosmic-ray source distribution, the height of the cosmic ray propagation halo and the HI spin temperature. This is important since the standard diffuse emission models distributed for LAT data analysis are optimized for analysis of point sources and compact extended sources \citep{2016ApJS..223...26A}, and the anlysis must be validated using different models. The estimation of the systematic errors was done as in \cite{2016ApJS..224....8A} by comparing the source parameters obtained with each alternative model and those with the standard model. The significant detection of G350.6-4.7 (above 15$\sigma$ for all alternative models considered), gives confidence that the emission is indeed associated to a new extended high energy source.

Systematic errors related to the imperfect knowledge of the effective area were estimated following the LAT team recommendations\footnote{See https://fermi.gsfc.nasa.gov/ssc/data/analysis/scitools\\/Aeff\_Systematics.html}. Since this uncertainty and the resulting uncertainty from the background model are independent, the total systematic error was obtained by adding them in quadrature.

\section{Discussion}\label{sec:discussion}
It is reasonable to attribute the origin of the gamma-rays to the shell of a previously unknown SNR. If located at the distance to the pulsar PSR J1747-4036 (3.4 kpc) the object size would be around 200 pc and the center of the disc would be located around 250 pc from the plane of the Galaxy. SNRs with diameters of the order of a few degrees are known and a significant number have been seen as extended objects by the LAT \citep{2016ApJS..224....8A}. Well-known examples are the Cygnus Loop \citep{2011ApJ...741...44K} and RX J0852.0-4622 \citep{2011ApJ...740L..51T}. However, not many SNRs are known to have a diameter greater than 100 pc, although large objects could not have detected due to observational selection effects \citep{1991PASP..103..209G}. Dynamic considerations of the expansion and evolution of a SNR can give a physical justification for an upper limit in their size \citep[e.g.,][]{1988ApJ...334..252C,1993ApJ...417..187S}. It is possible then that if this new source is a SNR it is much closer than 3.4 kpc and hence unrelated to PSR J1747-4036. This pulsar is classified as a millisecond pulsar \citep[e.g.,][]{2013ApJS..208...17A}, and thus if related to the G350.6-4.7, the SNR would also be too old to show a GeV spectrum as the one found here \citep[see e.g.,][]{2016ApJS..224....8A}. The Cygnus Loop SNR is located at about 540 pc \citep{2005AJ....129.2268B} and has a diameter of $\sim35$ pc. If the same physical diameter is assumed for G350.6-4.7, the resulting distance would be similar. However, the LAT spectrum of the Cygnus Loop \citep{2011ApJ...741...44K} is quite different to that reported here for G350.6-4.7 so these two objects likely have very different properties. On the other hand, with a diameter of $\sim2\degr$ (or $\sim$26 pc), RX J0852.0-4622 shows a GeV spectrum that is similar in shape to that of G350.6-4.7 \citep{2011ApJ...740L..51T}. Both of these previously known SNRs have very clear counterparts at other wavelengths.

A recently discovered source of gamma-ray emission that shows very similar features to those of G350.6-4.7 at high energies is reported at the location of the SNR G150.3+4.5 in the 2FHL catalog \citep{0067-0049-222-1-5}. The gamma ray emission was described by a disc template of radius $1^{\circ}.27$ and it shows a hard spectrum (power law index $1.66\pm 0.20$). SNR G150.3+4.5, which is thought to be associated to this gamma-ray source, was itself recently discovered \citep{2014A&A...567A..59G}. It is possible that the gamma-rays from G350.6-4.7 are produced in the shell of a new SNR. There might also be a contribution to the gamma-ray emission from the recently discovered SNR G351.0-5.4. However, this SNR has a small size compared to the gamma-ray source G350.6-4.7 and it is likely old \citep{2014A&A...568A.107D}.

There are candidate SNR shells that are detected at TeV energies and have no known counterparts at other wavelengths, such as HESS J1912+101 \citep{2015arXiv150903872P}, located at the coordinates $(l,b)=44\degr\llap{.}39,-0\degr\llap{.}07$. This source has no known counterpart at GeV energies and one of the reasons could be that the source has a hard GeV spectrum that is below the LAT sensitivity. It is possible that \ \ \ \ \ \ \ \ \ \ \ \ \ \ \ \ \ \ \ G350.6-4.7 is an example of this kind of TeV shells but for which the LAT telescope is able to detect the hard GeV spectrum. Follow up radio, X-ray and particularly TeV observations of G350.6-4.7 are necessary to understand its nature.

\begin{figure}
    \includegraphics[width=\columnwidth]{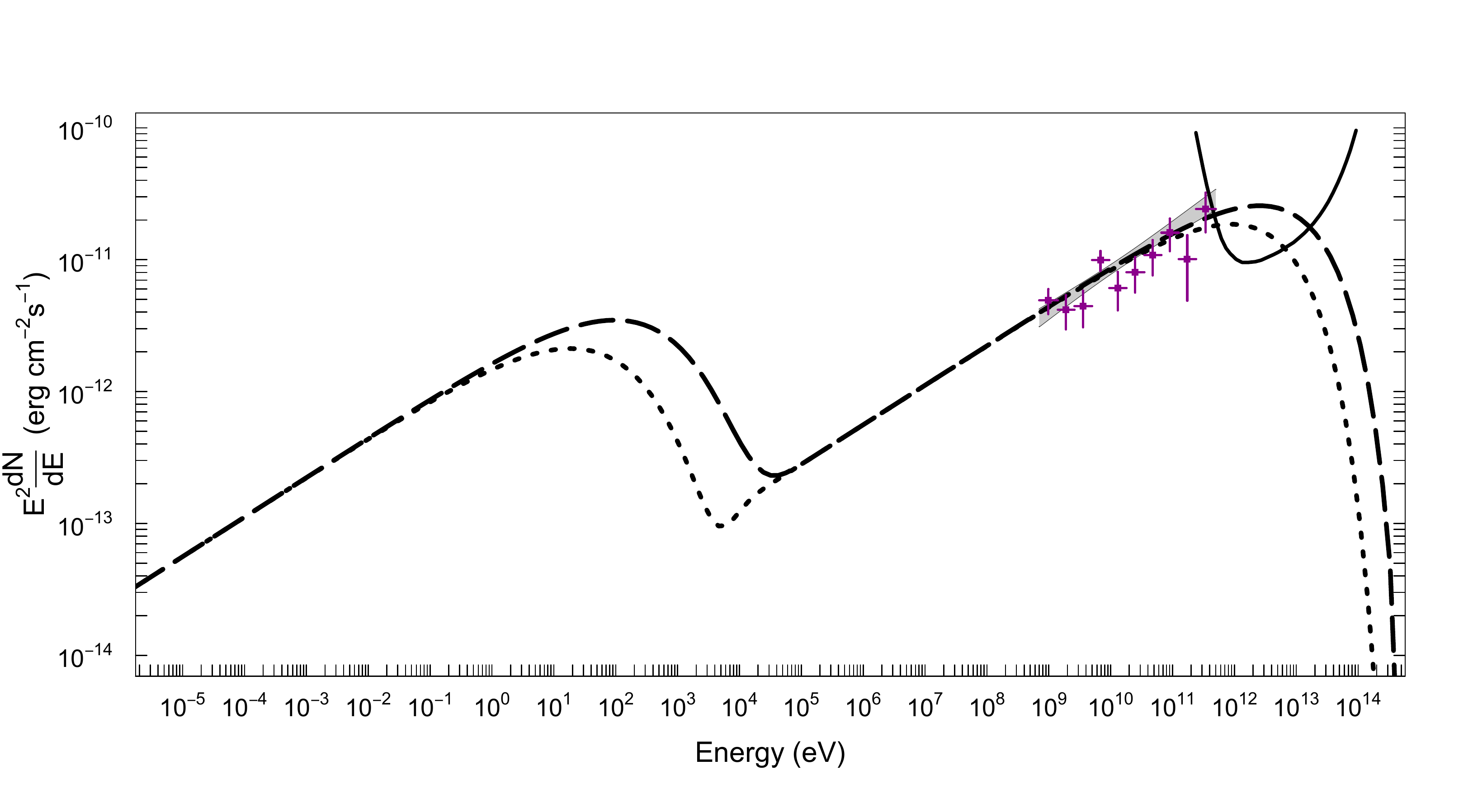}
    \caption{The spectral energy distribution of G350.6-4.7 is shown here above 0.7 GeV. The squares are the LAT data points with statistical errors and the shaded region represents the 1$\sigma$ fit uncertainty. Two versions of a simple leptonic model with electron cutoff energies of 35 TeV (dotted line) and 80 TeV (dashed line) are also shown. The solid line is the approximate H.E.S.S. sensitivity for a 100-hr observation of an extended source as explained in the text.}
    \label{fig2}
\end{figure}

It might be natural to assume from the spectral shape of G350.6-4.7 that the gamma-rays are produced by IC scattering of high-energy electrons. Fig. \ref{fig2} shows the SED of the source with a leptonic model calculated with the naima modeling package \citep{2015arXiv150903319Z}. The particle distribution is a power-law with a cutoff in energy ($\epsilon$) of the form $\frac{dN_e}{d\epsilon}\propto \epsilon^{-s}\cdot \mbox{e}^{-\epsilon/\epsilon_c}$. The particle cutoff energy is of course not constrained by the LAT data. Two models resulting for $\epsilon_c=35$ TeV and $\epsilon_c=80$ TeV, with the same particle spectral index $s=2.4$ and total energy content of \ \ \ \ \ \ \ \ \ \ \ \ \ \ \ \ \ \ \ \ \ $1.3\times10^{49}$ erg, are shown with the SED. The particles are assumed to be uniformly distributed in a spherical volume of radius 15 pc located at a distance of 500 pc from Earth. Since the true location of G350.6-4.7 is unknown, for simpicity, the photon field used in the model to serve as seed for upscattering is the Cosmic Microwave Background (CMB). The value for the magnetic field in the model is 1 $\mu$G, which is a typical value in the Galaxy. Based on the source size and the other parameters, the resulting electron energy density at the source would be 19 eV/cm$^{3}$.

For $\epsilon_c=35$ TeV, the peak of the synchrotron spectral energy distribution is seen at a photon energy of $20$ eV. The IC cooling time of a high-energy lepton with Lorentz factor $\gamma=7\times10^7$ (which corresponds to a particle energy \ \ \ \ \ \ \ \ \ \ \ \ \  $\epsilon \sim 35$ TeV) interacting with the CMB is \citep{1994hea2.book.....L}

\begin{equation}
  \tau = \frac{\epsilon}{\frac{4}{3}\sigma_T c \gamma^2 U_{\mbox{\tiny CMB}}} = \frac{2.3\times10^{12}\,\,\mbox{yr}}{\gamma} \approx 33 \,\,\mbox{kyr.}
  \label{ictime}
\end{equation}

If the source age is not higher than this value it is likely that it could be detected at TeV energies for the scenarios shown in Fig. \ref{fig2}. The approximate sensitivity for a 100-hr observation by H.E.S.S., which was taken from the point source sensitivity plot in \cite{2013APh....43..348F} and degraded by a factor of 10 to account for the fact that the source is extended as indicated by the observatory external proposal guidelines\footnote{https://www.mpi-hd.mpg.de/hfm/HESS/pages/home/proposals/}, can be seen in the SED plot also.

Another scenario for the origin of the gamma-rays is hadronic emission from cosmic rays accelerated in the shell of a SNR. The hadronic gamma ray flux above a photon energy of 1 GeV from a SNR located at a distance $d$ by a population of cosmic rays with a differential energy spectrum of the form $\epsilon^{-2.1}$ is given by \citep{1994A&A...287..959D}

\begin{equation}
    F \approx 1.8\times 10^{-7}\,\theta\, \left(\frac{E_{\mbox{\tiny SN}}}{10^{51} \,\mbox{erg}}\right)\,\left(\frac{d}{1\,\mbox{kpc}}\right)^{-2} \, \left(\frac{n}{1 \,\mbox{cm}^{-3}}\right)\,  \,\,\mbox{cm}^{-2}\,\mbox{s}^{-1}.
  \label{pp}
\end{equation} Here, $\theta$ is the fraction of the total supernova explosion energy ($E_{\mbox{\tiny SN}}=10^{51}$ erg) converted to cosmic ray energy and it is taken as 0.1 as usual \citep[e.g.,][]{2013A&A...553A..34D}, and $n$ is the target gas density. Using these values and the measured flux this translates to $n_1/d_{1}^2 \approx 0.2$, where the distance $d_1$ is in units of 1 kpc and $n_1$ is in units of 1 cm$^{-3}$. For $d_1=0.5$ this implies $n_1=0.05$.

Galactic gas distribution models and constraints on the cosmic ray energetics for off-plane gamma ray sources such as HESS J1507-622 \citep{2011AdSpR..47..640D}, observed 3$\degr\llap{.}$5 away from the plane of the Galaxy, predict an ambient density of $n_1\sim 0.55$ for a source distance of 500 pc. Using this density value and distance, the hadronic scenario for \ \ \ \ \ \ \ \ \ \ \ \ \ \ \ \ G350.6-4.7 requires a relatively low cosmic ray energy fraction of $\theta = 0.01$. On the other hand, a value closer to $\theta \sim 0.1$ would be allowed if the ambient density at the source is much lower than the one predicted by \cite{2011AdSpR..47..640D}, or the source distance is $\sim 1.6$ kpc. But for a higher distance of 2 kpc and $\theta=0.1$ the required density is $n_1\sim 0.8$ which might be high for the corresponding off-plane distance of \ \ \ \ \ \ \ \ \ \ \ \ \ 160 pc \citep{2011AdSpR..47..640D}. In a hadronic scenario, the source distance is then likely < 2 kpc, although this of course would change for a more energetic progenitor SN event or a different cosmic-ray acceleration efficiency.

The 70 Month Swift-BAT All-sky Hard X-Ray Survey \citep{2013ApJS..207...19B}, with a flux limit of \ \ \ \ \ \ \ \ \ \ \ \ \ \ \ \ \ \ \ \ \ \ $1.3\times10^{-11}$ erg s$^{-1}$ cm$^{-2}$ over 90\% of the sky, reveals several X-ray sources classified as low-mass binaries in the region. This flux upper limit is not costraining for the models discussed here. A PWN has a characteristic spectral energy distribution with two humps, one in X-rays attributed to synchrotron radiation of very-high energy leptons and another in the gamma-ray range produced by IC scattering of ambient photon fields by energetic leptons \citep[e.g.,][]{2009ApJ...703.2051G}. Assuming similar energetics and properties to those of a relatively extended PWN at GeV energies, Vela X \citep{2013ApJ...774..110G}, which is 290 pc away \citep{2001ApJ...561..930C}, the source G350.6-4.7 would be closer or physically larger, but the LAT source associated to the Vela X PWN shows a relatively soft spectrum (unlike that of G350.6-4.7) and bright radio emission \citep{2013ApJ...774..110G}. For a similar or shorter distance, the radio emission from the PWN in G350.6-4.7 would be brighter but has not been seen. Thus the PWN scenario is less likely for G350.6-4.7.

\begin{figure}
    \includegraphics[width=\columnwidth]{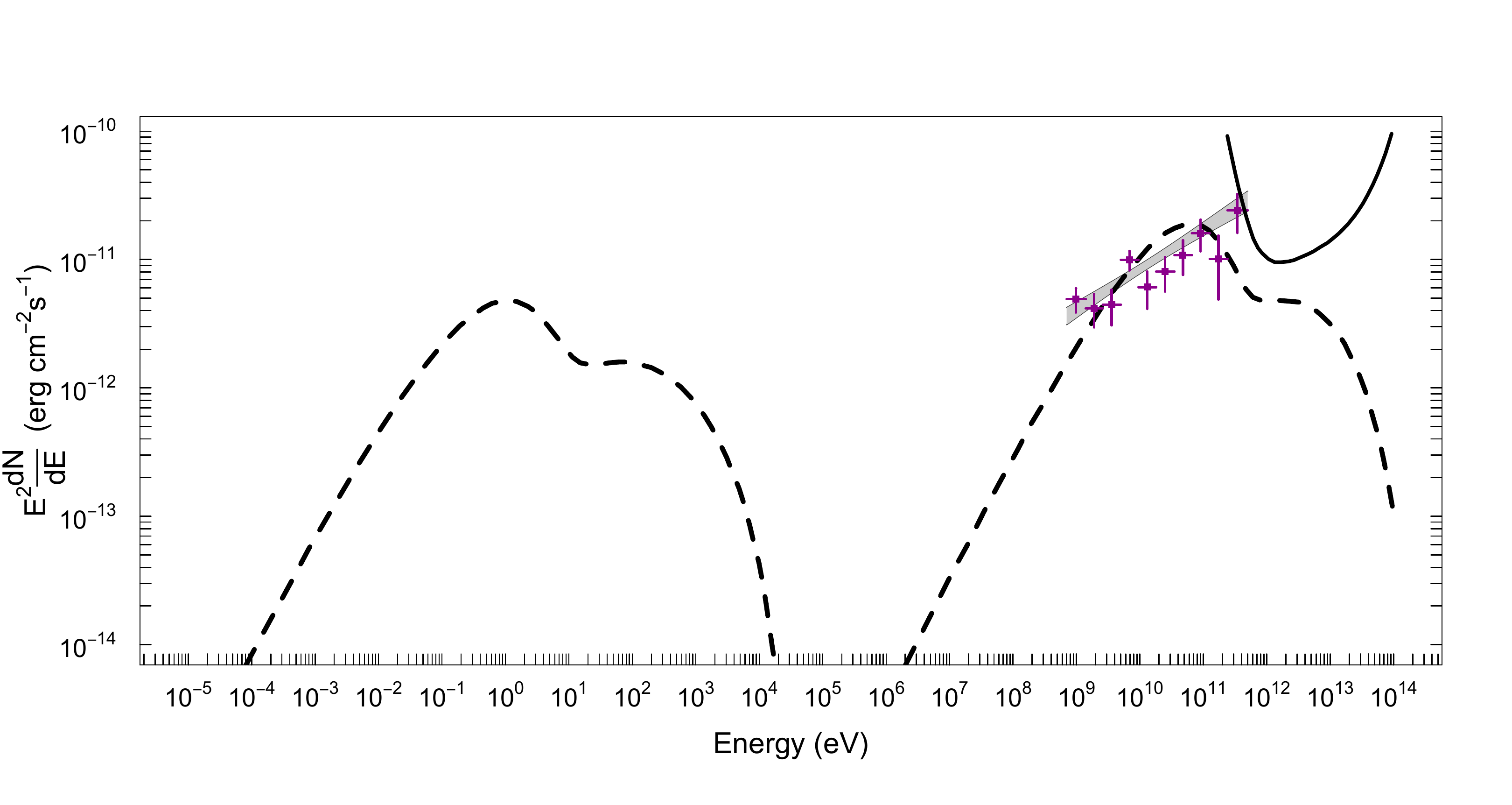}
    \caption{The same data shown in Fig. \ref{fig2} with a scaled-up relic PWN model (dashed lines) for the source HESS J1507-622.}
    \label{fig3}
\end{figure}

If there is no bright X-ray counterpart for G350.6-4.7, a possibility for the origin of the source is the IC scattering of energetic leptons in a relic PWN. The X-ray emission of evolved (relic) PWN is very low or absent leaving a VHE source with no counterpart. It is expected that the older a PWN becomes the lower its X-ray fluxes are, and the peak of the synchrotron hump is displaced to lower and lower energies while keeping the power-law component in the GeV band hard \citep{2012arXiv1202.1455M}. The details of a PWN evolution are expected to depend on the associated pulsar and the magnetic field, but are not well known. A typical prediction for the spectral energy distribution of an evolved  24 kyr-PWN taken from \cite{2013ApJ...773..139V} is shown in Fig. \ref{fig3} with the measured LAT spectral energy distribution of G350.6-4.7. The model was taken from that calculated for HESS J1507-622 in an ancient PWN scenario and it is scaled-up by a factor of 1.5. HESS J1507-622 is a peculiar source discovered in a H.E.S.S. Galactic Plane Survey \citep{2011A&A...525A..45H}, located relatively far from the Galactic plane ($\sim3\degr\llap{.}5$) and with no clear X-ray counterpart. However, for this model to be valid, G350.6-4.7 would have to be relatively close due to its large extent. It should be kept in mind that the relic PWN scenario is not well tested and HESS J1507-622 remains after all unidentified. Generally speaking, this is a plausible scenario for the origin of the high-energy photons if indeed there is no source counterpart at X-ray energies, although the particular model shown in Fig. \ref{fig3} fails to reproduce the shape of the spectral energy distribution. On the other hand there are several established SNRs showing gamma-rays of hadronic and leptonic origin. It is not yet known if there is a counterpart for G350.6-4.7 at lower energies but studies in the future might confirm the SNR origin of the emission.

Finally, the region around G350.6-4.7 is located within the Fermi Bubbles, a pair of very large structures seen in gamma-rays by the \emph{Fermi} satellite \citep{2010ApJ...724.1044S}. The spectrum of the bubbles above 1 GeV is described by a power-law with an exponential cutoff with index $\sim1.87$ and cutoff energy of $\sim 113$ GeV, showing no spectral variations across the bubbles \citep{2014ApJ...793...64A}. It was shown in Section \ref{subsec:spec} that the spectrum of G350.6-4.7 is different and thus a possible relation between them is unlikely.

Extended gamma-ray sources such as G350.6-4.7 offer an exciting possibility to carry out spatially-resolved spectral modeling with the LAT. This is relevant to understand particle diffusion in some sources interacting with ambient material as well as to probe the local acceleration properties of a source. It becomes important to constrain its distance. The necessary future studies to unveil the nature of G350.6-4.7 will surely benefit from observations at other wavelengths.

\section*{Acknowledgements}
The detailed comments on the work made by different anonymous referees are highly appreciated and helped improve the clarity and quality of this paper. Financial support from Universidad de Costa Rica is acknowledged. This research has made use of NASA's Astrophysical Data System.



\bibliographystyle{mnras}
\bibliography{references}

%\begin{thebibliography}{99}
%\bibitem[\protect\citeauthoryear{Abdo et al.}{2010a}]{Abdo2010a}
%Abdo, A. A. et al., 2010a, ApJ, 709, L152
%\end{thebibliography}

%%%%%%%%%%%%%%%%%%%%%%%%%%%%%%%%%%%%%%%%%%%%%%%%%%

%%%%%%%%%%%%%%%%% APPENDICES %%%%%%%%%%%%%%%%%%%%%

%\appendix



% Don't change these lines
\bsp	% typesetting comment
\label{lastpage}
\end{document}